\documentclass[prA,showpacs,twocolumn,aps]{revtex4}
\usepackage{epsfig}
\usepackage{graphicx}
\usepackage{amssymb}

\begin{document}

\title{Autler-Townes spectroscopy of the $5S_{1/2}-5P_{3/2}-44D$ 
cascade of cold $^{85}$Rb atoms}
\author{B. K. Teo, D. Feldbaum, T. Cubel, J. R. Guest, P. R. Berman and 
G. Raithel}
\affiliation{FOCUS Center and Physics Department, University of Michigan, 
Ann Arbor,
MI48109-1120}
\date{\today}

\begin{abstract}
We study nonlinear optical effects in the laser excitation of Rydberg
states. $5S_{1/2}$ and $5P_{3/2}$ levels of $^{85}$Rb are
coupled by a strong laser field and probed by a weak laser tuned to 
the $5P_{3/2} - 44D$ Rydberg resonance. We observe high contrast 
Autler-Townes spectra which are dependent 
on the pump polarization, intensity and detuning. The observed behavior 
agrees with calculations, which include the effect of optical pumping.
\end{abstract}

\maketitle

% o.k.

% body of paper here

\begin{figure}[hbt]
\epsfig{figure=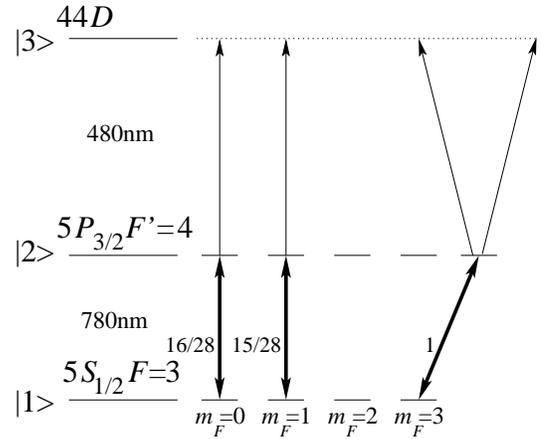,width=70mm}
\caption{Excitation scheme of $^{85}$Rb from the 5$S_{1/2}$ ground state 
to the 44$D$ Rydberg states using linear and $\sigma^+$ pump fields. The 
squares of the Clebsch-Gordon coefficients associated with the 
transitions are indicated next to the lower arrows. The probe field is 
linearly polarized in direction parallel to the linear pump field.} 
\label{level} \end{figure}

With the birth of laser spectroscopy in the late 60s, new classes of
nonlinear optical phenomena could be explored. One such measurement is
the optical analogue of the Autler-Townes splitting \cite{autler},
observed originally in the microwave domain. Following the work of 
Toschek and coworkers \cite{toschek69,toschek75a,toschek75b}
there have been a number of studies of
the Autler-Townes splitting in probe absorption, when a strong laser pump 
field drives a coupled transition in vapor cells or atomic beams
\cite{picque,vetter,liao,gray,delsart,sandeman}. The introduction of 
laser cooling techniques such as the
Magneto-Optical Trap (MOT) for neutral atoms has afforded precision
measurements and optical spectroscopy not limited by Doppler
effects or transit time broadening. Using frequency stabilized diode 
lasers, we extend previous Autler-Townes spectroscopy in MOTs 
\cite{atMOT1,atMOT2,atMOT3} to a new domain by using a high-lying Rydberg 
state with the excitation scheme shown in Fig.~\ref{level}. As a 
consequence, the decay rate associated with the final level is all but 
eliminated. This work represents a step towards driving ground 
state-Rydberg transitions, which has been proposed as the 
working element in quantum information storage 
schemes \cite{dipblock1,dipblock2}.

%The observed 
%lineshapes are in excellent agreement with calculations, which include 
%the effects of optical pumping. 

The measurement is performed at 60~Hz inside a vapor cell MOT with a 
background pressure of $10^{-10}$~Torr. In each experimental cycle, the 
MOT lasers are turned off for 150~$\mu$s and the pump and probe fields of 
duration 10~$\mu$s are switched on simultaneously. The pump couples 
the 5$S_{1/2}F=3$ ground state and the 5$P_{3/2}F^{\prime }=4$ 
intermediate state ($\lambda \approx780$~nm, saturation intensity 
$I_{\rm{sat}}$=1.64~mW/cm$^{2}$ and decay rate $\gamma_{2}/2\pi 
$=5.98~MHz,), while a linearly polarized probe field ($\lambda 
\approx480$~nm) weakly couples the intermediate state to the 44$D$ 
Rydberg states. The pump beam is 
collimated to a FWHM of 2.2~mm after spatial filtering 
through a single mode optical fiber. The probe beam is focused to a size 
of 30~$\mu $m and aligned anti-parallel to the center of the pump beam. 
In this arrangement, the pump field intensity is approximately constant 
throughout the probe volume, and field inhomogeneity effects are 
minimized. The Zeeman shifts arising from the MOT magnetic field 
(gradient of 10~G/cm) are less than 2~MHz in the excitation volume, which 
is below the spectroscopic resolution. The probe laser is a frequency 
doubled external cavity diode laser locked to a temperature regulated and 
pressure tunable Fabry-Perot cavity. The Rydberg population (lifetime 
$\approx60\mu$s) is monitored by counting electrons (gate = 100~$\mu$s) 
which originate from thermal ionization of Rydberg states with a 
Multi-Channel Plate (MCP) detector. The detection efficiency is estimated 
to be around 1.5~\%.

\begin{figure}[hbt]
\epsfig{figure=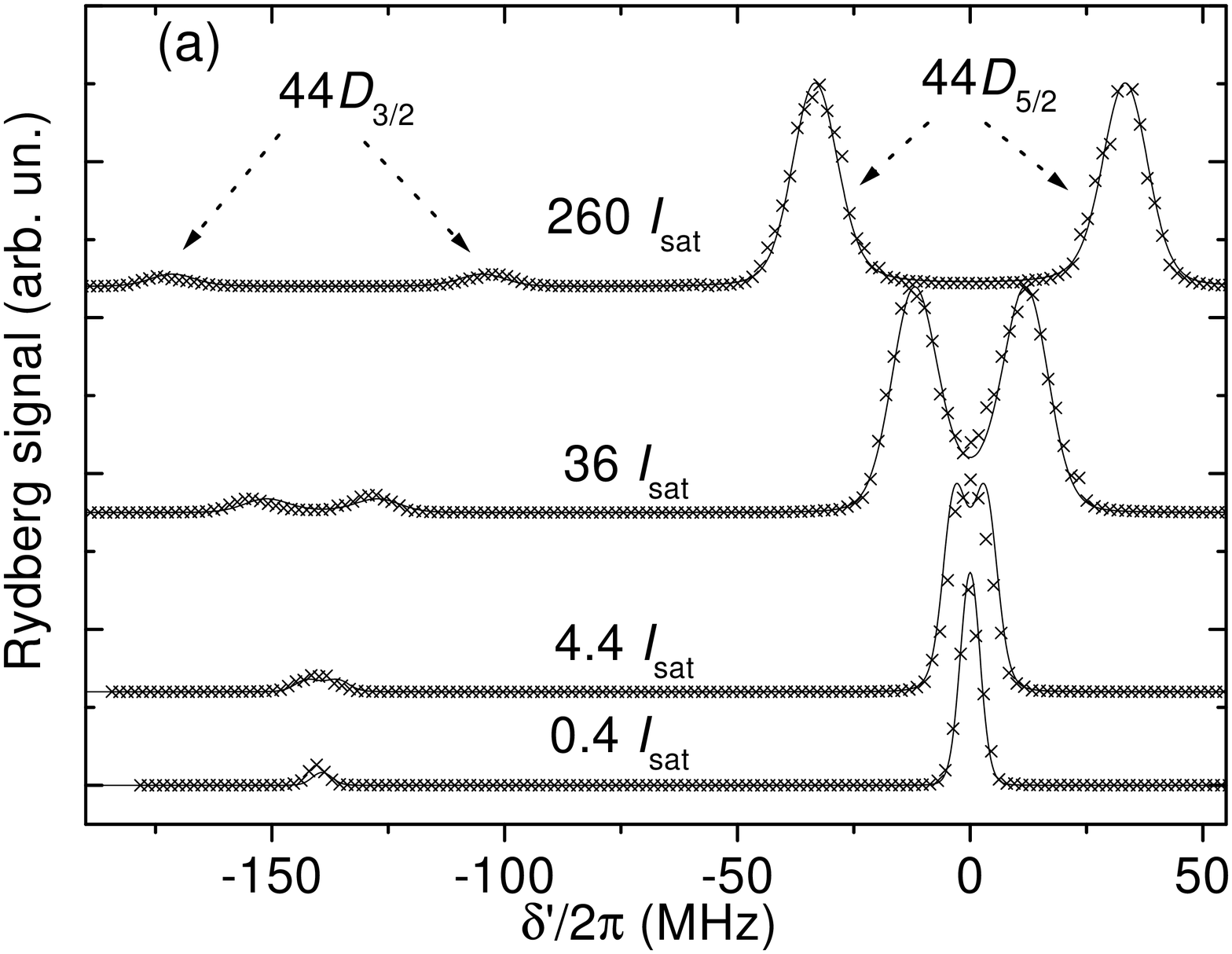,width=85mm} 
\epsfig{figure=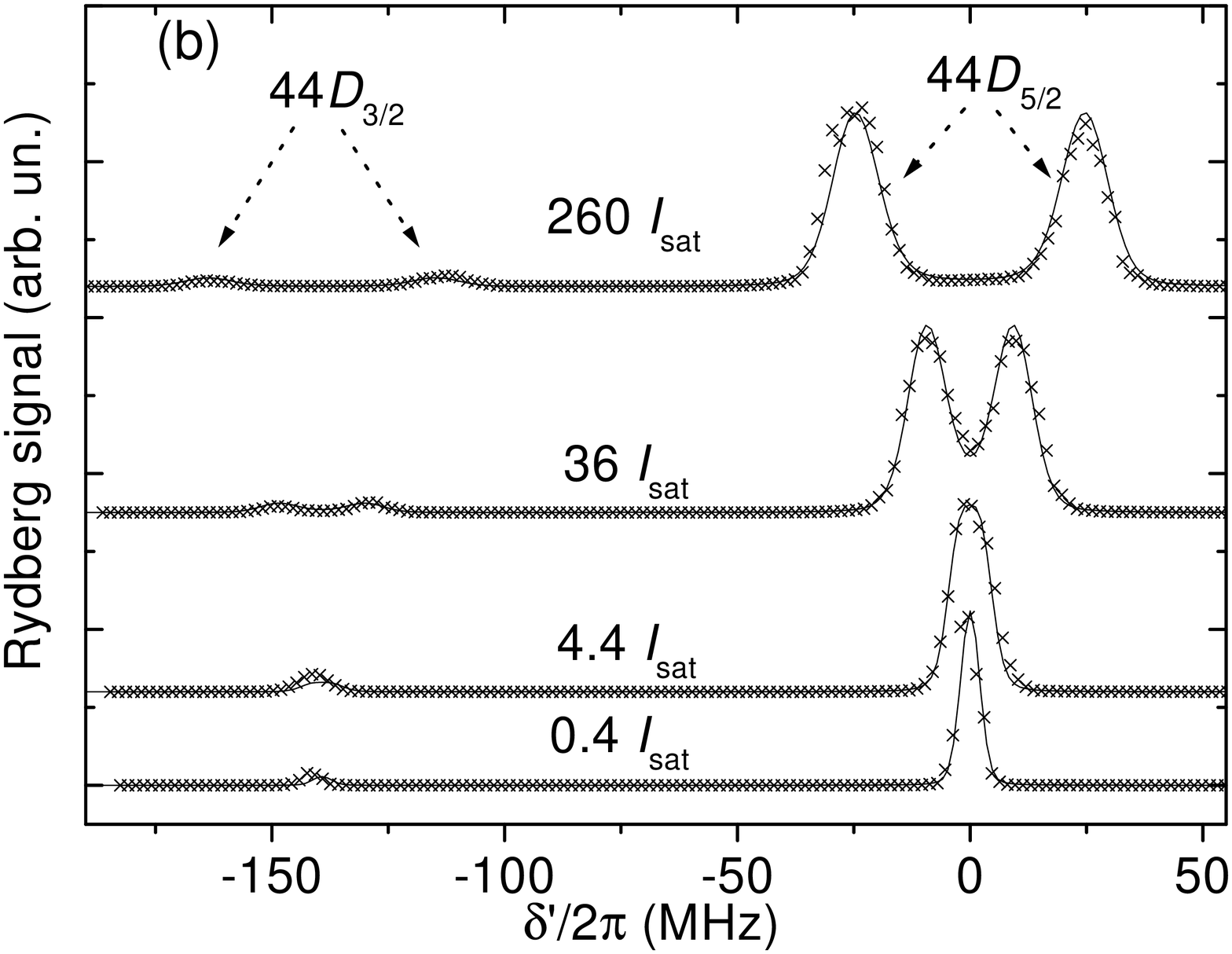,width=85mm}
\caption{Measured Autler-Townes spectra ($\times$) for zero pump 
detuning. The 44$D_{3/2}$ and the 44$D_{5/2}$ lines are shown for 
pump intensities of 260$I_{\rm{sat}}$, 36$I_{\rm{sat}}$, 
4.4$I_{\rm{sat}}$ and 0.4$I_{\rm{sat}}$ for $\sigma^+$ (a) and
linear (b)  pump polarizations. Solid lines are calculated lineshapes 
for intensities that have been adjusted by less than 10\% from the 
measured intensities.} \label{data}
\end{figure}

\begin{figure}[hbt]
\epsfig{figure=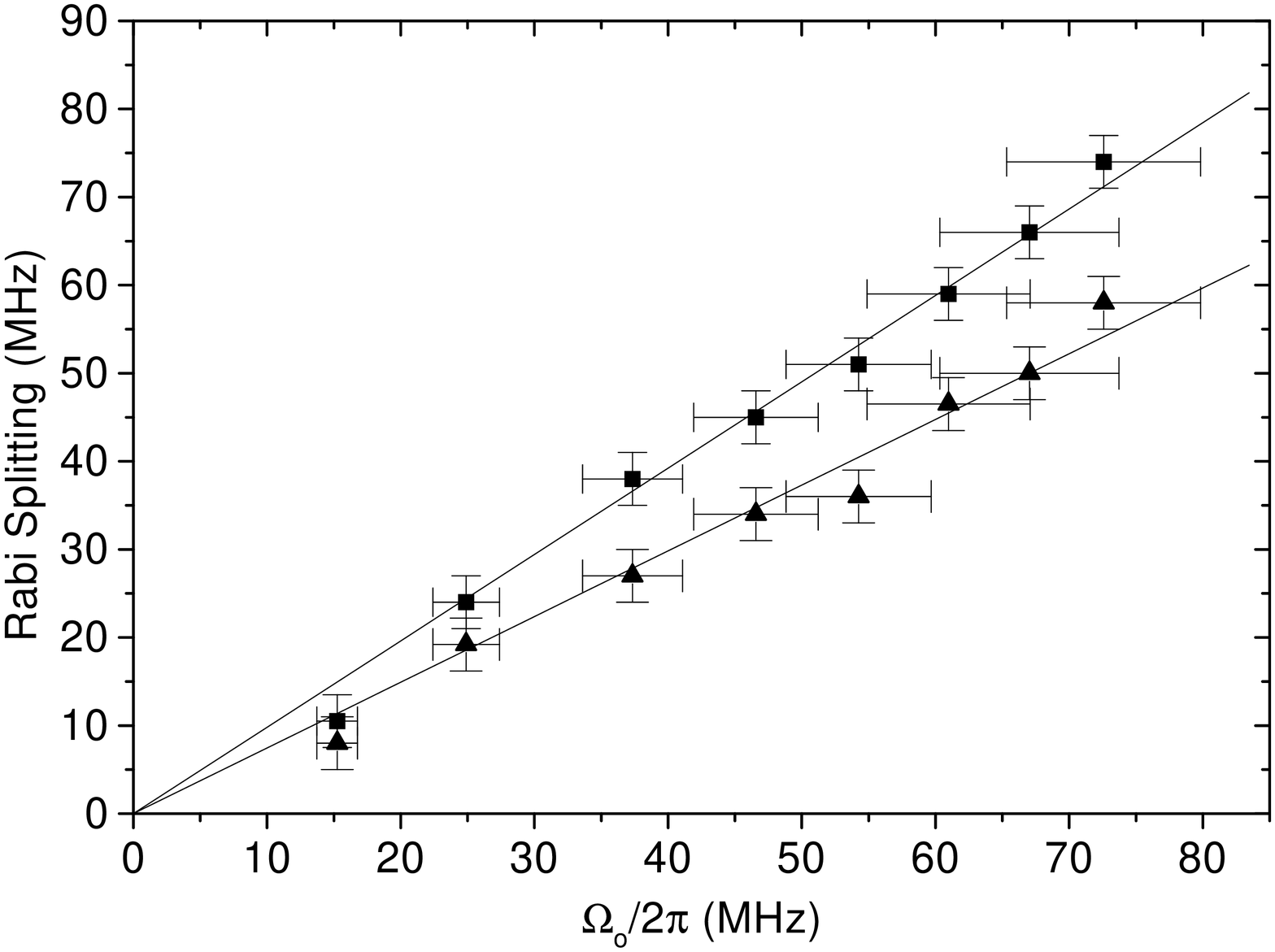,width=85mm}
\caption{Measured Rabi splitting at zero pump detuning for the 
44$D_{5/2}$ peaks for $\sigma^+$ ($\blacksquare$) and linear 
($\blacktriangle$) pump light. Solid lines-theory.} 
\label{fit}
\end{figure}

First, we consider the case of a resonant pump field (detuning, 
$\delta=\omega_{\rm{pump}}-\omega_{\rm{21}}=0$ where 
$\omega_{\rm{pump}}$ is the pump frequency and $\omega_{\rm{21}}$ is 
the transition frequency from $|1\rangle \rightarrow |2\rangle$). For a 
circularly polarized pump beam, the atoms are optically pumped into the 
$m_{F}=3$ ground state sublevel. Therefore, in steady state, the pump 
 field drives only a single transition ($F=3,m_{F}=3 \rightarrow
F^{\prime}=4,m_{F^{\prime}}=4)$ having a Rabi frequency $\Omega 
_{c}=\Omega
_{0}=\gamma _{2}\sqrt{\frac{s}{2}},$ where $s=I/I_{\rm{sat}}\ $is the
ratio of the pump intensity to the saturation intensity. 
For a pump field polarized linearly and parallel to the probe 
field's polarization, there are four 
contributions to the spectra originating from states differing in their 
values of $\left\vert m_{F}\right\vert $ (see below). However, the 
dominant contribution to the lineshapes comes from states having 
$m_{F}=0,\pm 1$. Since the Rabi
frequencies for these states, defined by 
\begin{eqnarray} 
\Omega _{\ell}(m_{F})=\langle F=3,m_{F};1,0|F^{\prime }=4,m_{F^{\prime 
}}\rangle\Omega _{0}, 
\end{eqnarray} 
are approximately equal, the Rabi splitting of the
different components is not resolved and the average Rabi splitting is 
$\Omega_{\ell}\approx0.74\Omega _{0}$. 
In Fig.~\ref{data}, the spectra as a function of probe 
detuning $\delta^{\prime}=\omega_{\rm{probe}}-\omega_{32}$ are shown, 
where $\omega_{\rm{probe}}$ is the probe frequency and $\omega_{32}$ is 
the 
transition frequency from $|2\rangle \rightarrow |3\rangle$. At low 
pump intensities, we see two resonances corresponding to the 44$D_{3/2}$ 
and 44$D_{5/2}$ lines which are separated by 140~MHz \cite{calibration}. 
The FWHM of these spectra lines are $\approx$ 4.5~MHz and can 
be attributed to the laser line width. As the pump intensity increases, 
the spectral lines broaden and the Autler-Townes 
components eventually separate. As shown 
in Fig.~\ref{fit}, the measured splittings are in excellent agreement 
with the Rabi frequencies ($\Omega_c=\Omega_0$ and $\Omega_{\ell}=0.74 
\Omega_0$), with $\Omega_0$ calculated using the measured pump beam 
intensities.

\begin{figure}[hbt]
\epsfig{figure=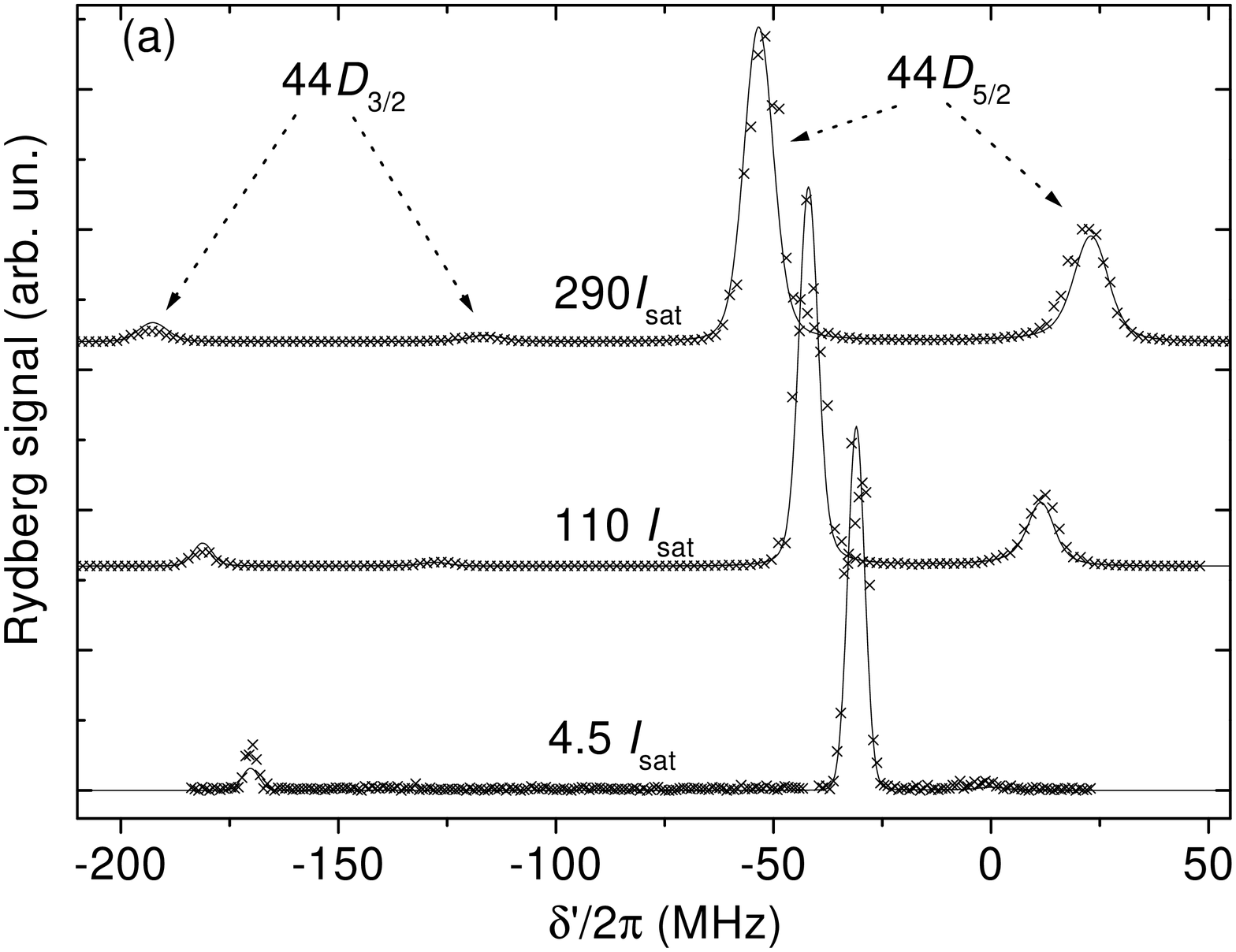,width=85mm} 
\epsfig{figure=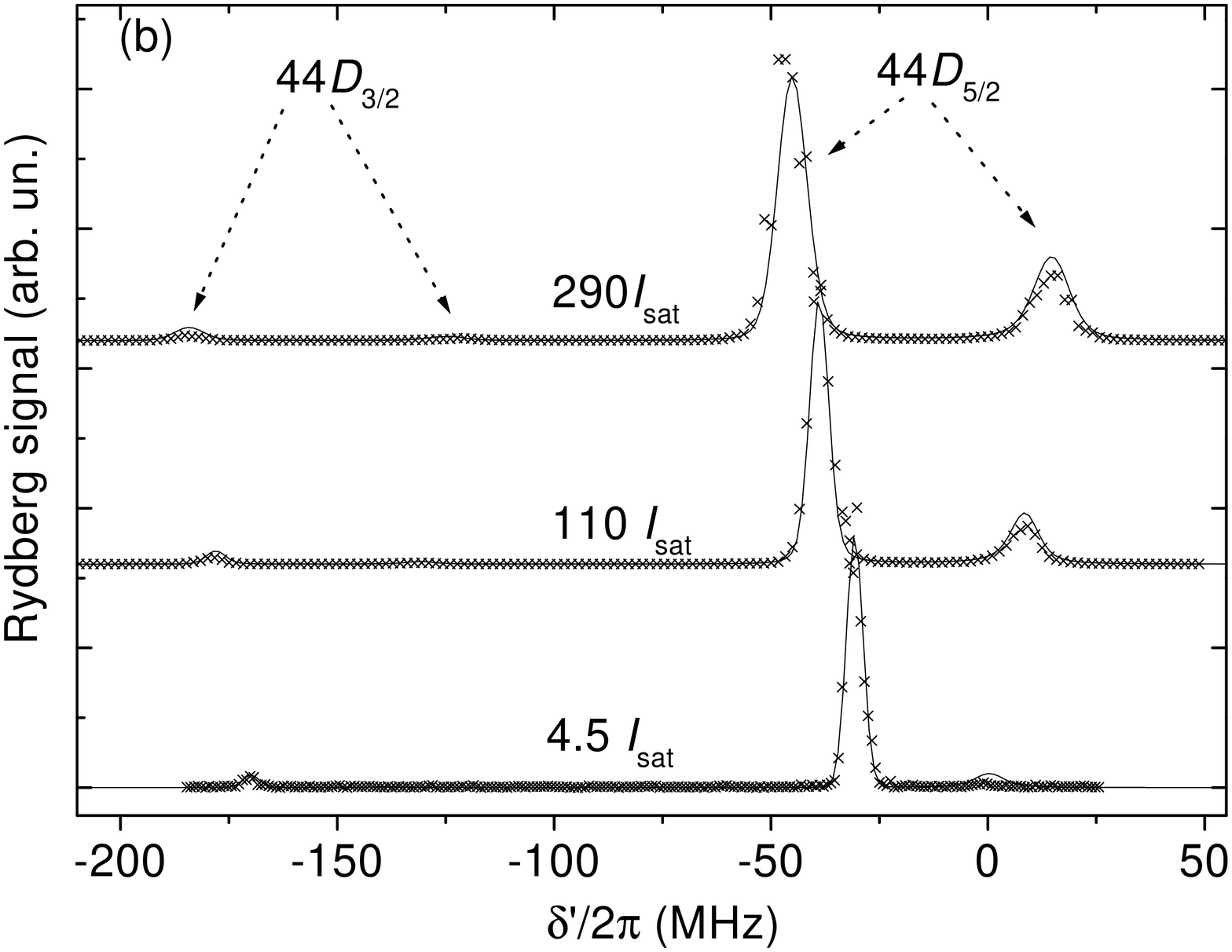,width=85mm}
\caption{Measured Autler-Townes spectra ($\times$) at 30~MHz (blue) pump 
detuning. The 44$D_{3/2}$ and the 44$D_{5/2}$ lines are shown for pump 
intensities of 290$I_{\rm{sat}}$, 110$I_{\rm{sat}}$, and 
4.5$I_{\rm{sat}}$ for $\sigma^+$ (a) and linear (b) pump polarizations. 
Solid lines are calculated lineshapes.}
\label{data30}
\end{figure}

\begin{figure}[t]
\epsfig{figure=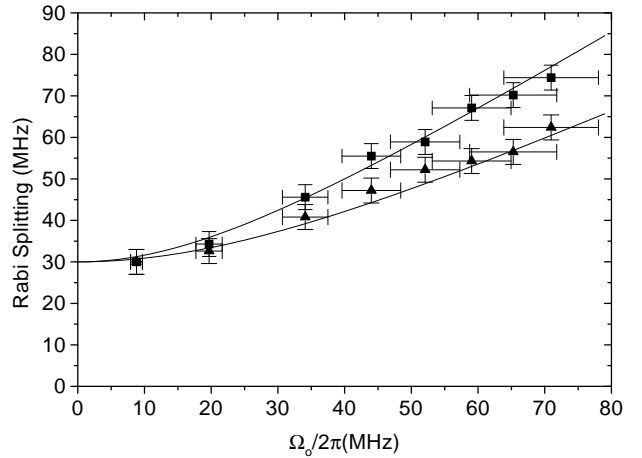,width=85mm}
\caption{Measured Rabi splittings at 30~MHz (blue) pump 
detuning for the 44$D_{5/2}$ peaks for $\sigma^+$ ($\blacksquare$) and 
linear ($\blacktriangle$) pump light. Solid lines-theory.}
\label{fit30}
\end{figure}

With the pump beam detuned to the blue by $\delta/2\pi=30$~MHz, the two
Autler-Townes components now have different amplitudes 
as shown in Fig.~\ref{data30}. The stronger 
component is red-detuned with respect to $\omega_{32}$ and corresponds to 
a two photon excitation process, while the weaker component corresponds 
to an off-resonant stepwise excitation. 
The splitting of the Autler-Townes peaks is
given by the generalized Rabi frequency, 
$\sqrt{\Omega_{c}^2+\delta^2}$ and $\sqrt{\Omega_{l}^2+\delta^2}$ for 
$\sigma^+$ and linear pump polarizations, 
respectively.
The measured Rabi splittings are in excellent agreement with the 
theoretical curves for both types of pump polarization, as shown in 
Fig.~\ref{fit30}. The splittings for the $\sigma^+$ case are larger than 
those observed in the linear case owing to the larger Clebsch-Gordon 
coefficient.

\begin{figure} [t]
\epsfig{figure=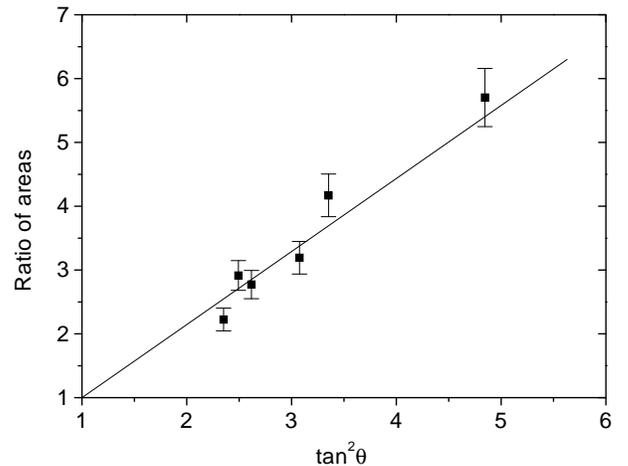,width=90mm}
\caption {Ratio of areas of the two 44$D_{5/2}$ Autler-Townes 
components for $\sigma^+$ pump light. Solid line-theory.} 
\label{ratio} 
\end{figure}

Theoretical expressions for the Autler-Townes spectra are derived
by solving the full set of density matrix equations to lowest order in the
probe field, but to all orders of the pump field. It is assumed that the
atom-field interaction time is sufficiently long for optical pumping to
establish a steady-state distribution. In the case of circularly polarized
pump field radiation, this implies that the pump field acts only between the 
$F=3,m_{F}=3$ and $F^{\prime}=4,m_{F^{\prime}}=4$ levels. The 
Autler-Townes signal is proportional to the total population in the 44$D$ 
manifold. Arbitrarily normalizing the signal to the $\sigma^+$ 
pump $D_{5/2}$ component 
(that is the integral over $\delta ^{\prime }$ of this component 
is set equal to unity), we take as our signal 
\begin{eqnarray}
S_{c}=\frac{1}{\pi }\sum\limits_{J=3/2,5/2}\frac{w_{c}(J)}{w_{c}(5/2)}Re\left[ \frac{\mu _{23}(J)}{\mu _{23}(J)\mu _{13}(J)+\left( \Omega
_{0}^{2}/4\right) }\right],
\end{eqnarray}
where $\mu _{23}=(\gamma _{2}+\gamma _{3})/2+i\delta ^{\prime }(J)$, $\mu
_{13}=\gamma _{3}/2+i[\delta +\delta ^{\prime }(J)]$, $\gamma 
_{3}\approx 16\times10^{3}$s$^{-1}$ is 
the decay rate of the 44$D$ levels \cite{gamma3}, $\gamma =\gamma _{2}/2$ 
and 
$w_{c}(J)$ is a weighting function 
equal to
\begin{widetext}
\begin{eqnarray}
w_{c}(J)=18\left\{ 
\begin{array}{ccc}
2 & 1 & 1 \\ 
3/2 & 1/2 & J%
\end{array}%
\right\} ^{2}(2J+1)\sum_{F}\left( \left\langle 4,4;1,-1|F,3\right\rangle
^{2}+\left\langle 4,4;1,1|F,5\right\rangle ^{2}\right) \left\{ 
\begin{array}{ccc}
J & 1 & 3/2 \\ 
4 & 5/2 & F%
\end{array}%
\right\} ^{2},
\end{eqnarray}
\end{widetext}
where $\left\{ \ldots \right\} $ is a 6-J symbol. 
From this expression it
follows that the ratio of the $J=$5/2 to 3/2 signal is 16.5. In the 
limit that
the generalized Rabi frequency $\Omega=\sqrt{\Omega_{0}^{2}+\delta^{2}}\gg\gamma$, it is possible to reexpress the lineshape using a dressed
atom approach \cite{salomaa} as 
\begin{widetext}
\begin{eqnarray}
S_{c}=\frac{1}{\pi}\sum\limits_{J=3/2,5/2}\frac{w_{c}(J)}{w_{c}(5/2)}\left[
\frac{\Gamma _{1}\cos ^{2}\theta }{\Gamma _{1}^{2}+\left( \delta ^{\prime
}(J)+\frac{\delta +\Omega }{2}\right) ^{2}}+\frac{\Gamma _{2}\sin ^{2}\theta 
}{\Gamma _{2}^{2}+\left( \delta ^{\prime }(J)+\frac{\delta -\Omega }{2}%
\right) ^{2}}\right] ,
\label{sc}
\end{eqnarray}
\end{widetext}
where $
\Gamma _{1}=\frac{\gamma _{3}}{2}\cos ^{2}\theta +\frac{(\gamma _{2}+\gamma
_{3})}{2}\sin ^{2}\theta;
\text{ \ }\Gamma _{2}=\frac{\gamma 
_{3}}{2}\sin
^{2}\theta +\frac{(\gamma _{2}+\gamma _{3})}{2}\cos ^{2}\theta$ and 
$\cos ^{2}\theta =\frac{1}{2}\left( 1+\frac{\delta }{\Omega }\right).$
In this form, it is clear that the ratio of the amplitude of the components
of the Autler-Townes doublet for a given $J$ is equal to $\tan ^{4}\theta $
and the ratio of the areas is equal to $\tan ^{2}\theta $. The ratio of 
the areas is independent of laser line width. In Fig.~\ref{ratio} we plot 
the ratio of the areas versus tan$^{2}\theta $ and see that experiment is 
in excellent agreement with theory.

For the case of a linearly polarized pump field, there are four 
contributions to each lineshape component from states differing in 
$\left\vert m_{F}\right\vert $. Using the normalization specified above, 
one finds the lineshape to be given by
\begin{widetext}
\begin{eqnarray}
S_{\ell }=\frac{1}{\pi }\frac{\gamma ^{2}(1+s)+\delta ^{2}}{\Omega _{0}^{2}/4%
}\sum\limits_{J=3/2,5/2}\sum\limits_{m}\frac{w_{\ell }(J,m)}{w_{c}(5/2)}%
\frac{\Omega _{\ell }(m)^{2}}{\gamma ^{2}(1+\frac{s}{2})+\delta ^{2}}Re\left[ \frac{\mu _{23}(J)}{\mu _{23}(J)\mu _{13}(J)+\left[ \Omega _{\ell
}(m)^{2}/4\right] }\right] \rho _{1m,1m}^{(0)}
\label{sigeqn}
\end{eqnarray}
where
\begin{eqnarray}
w_{\ell }(J,m)&=&18\left\{ 
\begin{array}{ccc}
2 & 1 & 1 \\ 
3/2 & 1/2 & J%
\end{array}%
\right\} ^{2}(2J+1) \sum_{F}\left\langle 
4,m;1,0|F,m\right\rangle 
^{2}\left\{ 
\begin{array}{ccc}
J & 1 & 3/2 \\ 
4 & 5/2 & F%
\end{array}%
\right\} ^{2}
\end{eqnarray}
\end{widetext}
and $\rho _{1m,1m}^{(0)}$ is the steady state ground state population of
sublevel $m$. Note that, in strong fields $\sum_{m}\rho _{1m,1m}^{(0)}\neq 1$%
, since there is non-negligible population in the $5P_{3/2}(F=4)$ level. It
is also possible to write an expression analogous to Eq.~(\ref{sc}) in a 
dressed basis by replacing $\left[ \frac{\mu _{23}(J)}{\mu _{23}(J)\mu 
_{13}(J)+\left[
\Omega _{\ell }(m)^{2}/4\right] }\right] $ in Eq.~(\ref{sigeqn}) with  
$\left[ \frac{%
\Gamma _{1}(m)\cos ^{2}\theta (m)}{\Gamma _{1}^{2}(m)+\left( \delta ^{\prime
}(J)+\frac{\delta +\Omega (m)}{2}\right) ^{2}}+\frac{\Gamma _{2}(m)\sin
^{2}\theta (m)}{\Gamma _{2}^{2}(m)+\left( \delta ^{\prime }(J)+\frac{\delta
-\Omega (m)}{2}\right) ^{2}}\right] $, where the dressed state angle $\theta 
(m)$, the dressed state decay rates $\Gamma_j(m)$, and the generalized 
Rabi frequency $\Omega(m)$ are calculated using $\Omega_{\ell
}(m)$ rather than  $\Omega_0$.

From the numerical solutions, we find that, for the range of parameters in
our experiments, $\rho _{10,10}^{(0)}$ varies between 0.2 and 0.38, $\rho
_{11,11}^{(0)}$ between 0.12 and 0.24, $\rho _{12,12}^{(0)}$ between 0.036
and 0.06, and $\rho _{13,13}^{(0)}$ between 0.0012 and 0.0044. Thus, the $
m_{F}=0,\pm 1$ states give the dominant contributions. The ratio of the 
$D_{5/2}$ to $D_{3/2}$ component is found to be equal to 21 to within a 
few percent (it varies slightly with field intensity and detuning). The 
Rabi frequencies are equal to [$\Omega_{\ell 
}(0)=\sqrt{\frac{16}{28}}\Omega _{0},$ 
$\Omega
_{\ell }(1)=\sqrt{\frac{15}{28}}\Omega_{0}$, $\Omega_{\ell}(2)= 
\sqrt{\frac{12}{28}}\Omega _{0},$ $\Omega_{\ell}(3)= 
\sqrt{\frac{7}{28}}\Omega_{0}$];
however since the $m=0,\pm 1$ substates dominate, the Rabi 
spitting is approximately equal to $\Omega 
_{e}=\frac{1}{2}[\Omega_l(0)+\Omega_l(1)]\approx 0.74\Omega _{0}$, and 
the ratio of the amplitude of the
components of the Autler-Townes doublet for a given $J$ is equal roughly 
to $\tan^{4}\theta_{e}$, where $\theta_{e}$ is calculated using the 
effective Rabi frequency $\Omega_{e}$.

Since the laser line width is comparable with or larger than the 
level line widths, it is necessary to convolute the lineshapes with the 
spectral profile of the laser. Assuming that the width of the lines 
observed at the lowest 
intensities in Fig.~\ref{data} and Fig.~\ref{data30} is attributable 
solely to laser line width, we find that the laser profile is described  
well by a Gaussian with FWHM of 4.5~MHz. This leads to excellent 
agreement between the experimental and theoretical curves at low pump 
intensities.
However, to fit some of the curves in 
Fig.~\ref{data} 
and Fig.~\ref{data30}, it was necessary to increase the convolution width 
up to 10~MHz with increasing pump field intensity. We attribute this to 
the fact that the pump and probe beams were not perfectly aligned; 
inhomogeneities in the pump laser field then result in a distribution of 
Rabi frequencies, broadening the lines. The deviation of the data from 
the fits in Fig.~\ref{data30} at low pump intensities can be attributed 
to incomplete optical pumping.

In summary, we have observed high-resolution Autler-Townes Rydberg
spectra of $^{85}$Rb. The measured dependence of the Rabi
splittings and line strengths on the pump field intensity,
polarization and detuning are in excellent agreement with theory.
In the near future, we plan to extend the off-resonant
Autler-Townes measurements examined in this paper towards a
pulsed, coherent two-photon excitation from the ground state into
well defined Rydberg states. It should be possible to achieve
two-photon Rabi oscillations between ground- and Rydberg states by
starting with the detuned low-intensity regime of Fig.~\ref{data30}. The 
Rabi frequency of the upper $5P_{3/2}$ $\rightarrow$ $nD$ transition
will have to be increased and the Rabi frequency of the lower
transition $5S_{1/2}$ $\rightarrow$ $5P_{3/2}$ reduced, such that
the population in the intermediate $5P$ state is minimized while a
large two-photon Rabi frequency is maintained. The realization of
Rabi oscillations or, equivalently, $\pi$- and $2\pi$-pulses
\cite{pipulse} to
coherently transfer atoms into and out of individual Rydberg
levels is of particular interest, because such operations are an
important element in fast quantum gates that have been proposed in
context with neutral-atom quantum computing
\cite{dipblock1,dipblock2}.

This work was supported in part by NSF Grant Nos. PHY-0114336, 
PHY-0098016, PHY-9875553 and the U.S. Army Research Office Grant 
No. DAAD19-00-1-0412. J. R. G. acknowledges support from the 
Chemical Sciences, Geosciences and Biosciences Division of the Office of 
Basic Energy Sciences, Office of Science, U.S. Department of Energy.


\begin{thebibliography}{99}

\bibitem{autler} S. H. Autler and C. H. Townes, Phys. Rev. \textbf{100}, 703
(1955).

\bibitem{toschek69} T. H\"ansch \textit{et al}, Z. Physik \textbf{226}, 
293 (1969).

\bibitem{toschek75a} A. Schabert, R. Keil and P. E. Toschek, 
Opt. Commun. \textbf{13}, 265 (1975).

\bibitem{toschek75b} A. Schabert, R. Keil and P. E. Toschek, 
Appl. Phys. \textbf{6}, 181 (1975).

\bibitem{picque} J. L. Picque and J. Pinard,
J. Phys. B \textbf{9}, L77 (1976).

\bibitem{vetter} Ph. Cahuzac and R. Vetter, Phys. Rev. A \textbf{14}, 270
(1976).

\bibitem{liao} J. E. Bjorkholm and P. F. Liao,
Opt. Commun. \textbf{21}, 132 (1977).

\bibitem{gray} H. R. Gray and C. R. Stroud,
Opt. Commun. \textbf{25}, 359 (1978).

\bibitem{delsart} C. Delsart, J. C. Keller and V. P. Kaftandjian,
J. Phys. (Paris) \textbf{42}, 529 (1981).

\bibitem{sandeman} P. T. H. Fisk, H.-A. Bachor and R. J. Sandeman, Phys.
Rev. A \textbf{33}, 2418 (1986).

\bibitem{atMOT1} R. W. Fox \textit{et al}, Opt. Lett. \textbf{18}, 1456 
(1993).

\bibitem{atMOT2} A. G. Sinclair, B. D. McDonald, E. Riis and G. Duxbury, 
Opt. Commun. \textbf{106}, 207 (1994).

\bibitem{atMOT3} J. H. Marquardt, H. G. Robinson and L. Hollberg, J. Opt. 
Soc. Am. B \textbf{13}, 1384 (1996). 

\bibitem{dipblock1} D. Jaksch \textit{et al}, Phys. Rev. Lett. \textbf{85},
2208 (2000).

\bibitem{dipblock2} M. D. Lukin \textit{et al}, Phys. Rev. Lett. \textbf{87}%
, 037901 (2001).

\bibitem{calibration} K. C. Harvey and B. P. Stoicheff, Phys. Rev. Lett. 
\textbf{38}, 537 (1977)

\bibitem{gamma3} In our 300K radiation field, the width of the final 
level is estimated to be $\approx 16\times10^{3}$s$^{-1}$, {\sl{i.e.}} 
slightly larger than the natural line width of 
10.9$\times10^{3}$s$^{-1}$, but still much less than all other relevant 
widths.

\bibitem{salomaa} P. R. Berman and R. Salomaa, Phys. Rev. A \textbf{25}, 
2667 (1982).

\bibitem{pipulse} These $\mu$s $\pi$ pulses are much longer than the ns 
pulses typically used in photon echo experiments. See, for example, A. 
Flusberg, R. Kachru, 
T. Mossberg and S. R. Hartmann, Phys. Rev. A \textbf{19}, 1607 (1979).

\end{thebibliography}
\end{document}